# Improvement of Heat-Treated Niobium Surface by *In-situ* Plasma Treatment Applied to Superconducting RF Resonator.

C. Boutelaa[1]†, S. Gruszka[1], J. Yemane[1], C. Cheney[1], T. Gerardin[1], E. Mistretta[1], J. Demailly[1], S. Guilet[4], P. Kolb[2], J. Keir[2], B. Mercier[1], N. Prud'homme[3], G. Sattonnay[1], R. Laxdal[2], D. Longuevergne[1]

[1]IJCLab, Paris Saclay University, Orsay, France
[2]TRIUMF, Vancouver, Canada
[3]ICMMO, Paris Saclay University, Orsay, France
[4]Institut des NanoSciences de Paris, Sorbonne University, Paris, France

†Corresponding author: Chahinez.boutelaa@ijclab.in2p3.fr

**Abstract:**

A specific heat treatment at 300°C, named medium-temperature baking (Mid-T baking) is applied to superconducting radio-frequency (SRF) accelerating cavities to improve the quality factor ($Q_0$) at medium accelerating fields (10-25 MV/m). However, surface contamination can lead to the degradation of $Q_0$ instead. Plasma-based surface treatment provides an effective approach to eliminate contaminants from the Niobium surface. In this study an *in-situ* plasma cleaning process with argon containing 10% $O_2$ was performed to remove hydrocarbons from the Niobium surface. The treatment was applied before and after a heat treatment at 500°C under ultra-high vacuum conditions (Mid-T baking). Changes in chemical speciation and oxide layer alteration induced by plasma processing were analyzed using *in-situ* X-ray photoelectron spectroscopy (XPS) and *ex-situ* scanning electron microscopy (SEM). The results show that plasma treatment after 1 hour modifies the composition of Niobium oxides, converting a $Nb_2O_5$ layer into $NbO_2$. Furthermore, an *in-situ* plasma treatment before Mid-T baking also called "Plasma Bake" helps reduce unstable oxides such as $Nb_xO$ and significantly increases the proportion of metallic Niobium at the surface. The Niobium sample treated by plasma Bake showed a 53% reduction in carbide formation compared to Mid-T alone. Moreover, the C1s component attributed to Nb–C bonds shifts toward lower binding energy, indicating the formation of a more metallic NbC phase. Whereas without plasma treatment, the higher binding energy component observed after Mid-T baking is consistent with $Nb_2C$. Finally, the plasma allowed to remove the NbC and $Nb_2C$ formation at the surface.

## Introduction:

The preparation of cavity inner surfaces has been one of the major challenges in superconducting radio-frequency (SRF) accelerator technology. Accelerator performance depends directly on the physical and chemical characteristics of the SRF cavity surface. In recent decades various surface treatment methods have been proposed and explored to push to higher $Q_0$ at high accelerating gradient $E_{acc}$ of SRF cavities. A high temperature annealing at 800°C followed by a medium temperature baking (Mid-T bake) at 300°C for 3h after re-oxidation, is a novel surface doping procedure, typically achieving $Q_0$ values larger than $1 \times 10^{10}$ at 2.0 K but limiting the maximal achievable accelerating gradient to 25 MV/m [4,5,6,7]. The performance improvements associated with the Mid-T bake treatments have been attributed to the dissolution of the native oxide layer and the subsequent diffusion of oxygen into the near surface, which provides a positive doping effect [9]. It has also been observed that Mid-T treatments are associated with carbide formation depending on the temperature of the treatment [3]. Typical mid-T treatments occur at 300°C while maximum carbon formation occurs at 500C [3]. A Mid-T bake at 500°C for 3 h under ultra-high vacuum (UHV) conditions was chosen for the study in order to exacerbate the surface related phenomenon we are interested in studying.

*In-situ* cleaning methods, like plasma processing, have gained more and more interest from the SRF community as it reduces the effort required to regain SRF cavity performance in a linac application. Numerous studies have been conducted on the application of this cleaning process [12,27]. However, very few studies have focused on the fundamental understanding of the interaction between the plasma and the niobium surfaces, particularly when the surfaces have been prepared using a medium-temperature heat treatment under ultra-high vacuum conditions. In this study, an *in-situ* plasma processing and characterization approach was developed using a plasma composed of 90% Ar and 10% $O_2$ to investigate the effects of plasma treatment performed either before or after heat treatment at 500 °C. The carbon contamination, niobium carbides (NbC, Nb2C) and oxide layer formation under plasma cleaning were analyzed using in-situ X-ray photoelectron spectroscopy (XPS) and *ex-situ* scanning electron microscopy (SEM). Several plasma conditions were studied and compared to better understand the improvements that plasma treatments can bring during heat treatment under UHV. Section I of this paper describes the experimental procedure for two different cases. Section II presents the results and discusses modifications in carbon contamination and Niobium oxide after a Mid-T bake using both procedures.

## 1. Experiment and methods

### *1.1 Experimental set-up*

The *in-situ* Plasma-XPS-Baking experiments were performed in the Vacuum and Surfaces Platform at IJCLab. The set-up is equipped to perform in-*situ* XPS studies during annealing under UHV, with a dedicated plasma chamber as shown schematically in figure 1. The system consists of two vacuum chambers isolated by a valve to minimize the impact of the plasma treatment (done at $10^{-2}$ mbar) on the UHV chamber (maintained below $10^{-9}$ mbar). The plasma chamber in figure 1(b), is equipped with a remote inductively coupled plasma (ICP) source (IBSS Group GV10x UHV DS Asher) and the base pressure of the plasma chamber is $10^{-7}$ mbar before injection of reactive gas. The analysis chamber includes a heating sample holder and an X-ray source for XPS measurements (see the section I.3 for more details). The temperature is measured directly on the sample using a type K thermocouple. The initial pressure of this chamber is around $10^{-10}$ mbar. A transfer rod allows the sample to be moved from one chamber to the other without exposure to air.

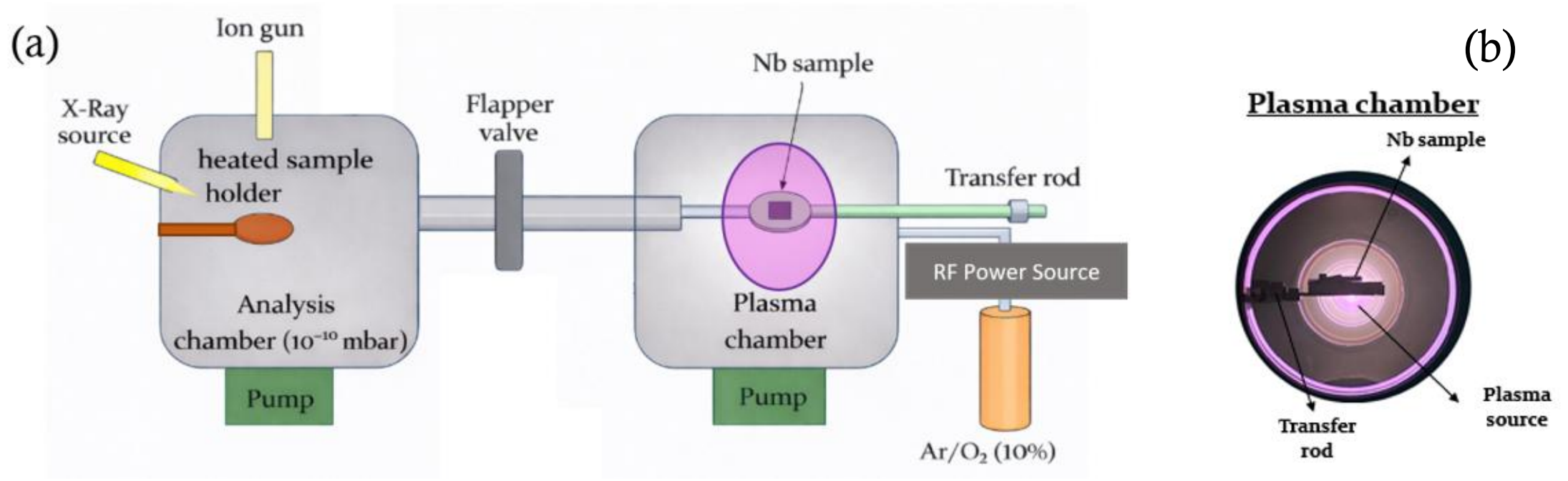

Figure. 1 : Experimental setup for plasma cleaning, XPS analysis, and sample annealing (a). Plasma chamber (b)

### *1.2 Sample preparation*

The bulk Niobium (Nb) used to fabricate SRF cavities comes in the form of metal sheets made from fine-grain niobium used in the TRIUMF linac with a residual resistance ratio (RRR) better than 250. At TRIUMF, the niobium was cut into 10 mm × 10 mm squares using a water jet. Each sample was then chemically treated in a buffered chemical polishing (BCP) solution (HF: $HNO_3$:$H_3PO_4$ = 1:1:2 by volume). The etch was conducted with the acid bath in a secondary vessel filled with an ice bath to keep the temperature blow 15°C which is typically the threshold for hydrogen absorption into the Nb. This process removed approximately 100 µm of surface material. The samples were then thoroughly rinsed with ultra-pure water. Thermal treatments of the Niobium samples were carried out in the TRIUMF induction furnace at 800 °C for 3 hours [3]. The heat treatment procedure consists of three stages: heating-up (ramp of 6.6°C/min), plateau at the defined temperature under UHV ($10^{-7}$ Torr), and natural cooling-down (about 2-3 hours to reach room temperature. Afterwards, each sample was subjected to flash BCP etching of about 10 µm followed by ultra-pure water rinsing. The samples were rinsed again at IJCLab with ultrapure water before being used in this experiment. Two Niobium samples were used for this study and followed two different paths (A and B). The two paths are outlined in figure 2.

In the first one (procedure A), the sample was analyzed by XPS (baseline). The Mid-T bake treatment, which we characterized as annealing at 500°C for 3 hours, was then applied to this sample. The heat treatment procedure consisted of three stages: heating-up at a rate of 4.1 °C/min for 2 hours, plateau at 500°C during 3h under UHV ($10^{-9}$ mbar), and natural cooling until room temperature for 2h. The initial CO and $CO_2$ partial pressures were extremely low (≈$10^{-10}$/$10^{-11}$ mbar) before heating. During annealing, outgassing from the sample increased the $CO_2$ partial pressure to approximately $10^{-10}$ mbar and the CO partial pressure to approximately $10^{-9}$ mbar. The sample was analyzed by XPS directly after the baking (Mid-T bake) without air exposure. Then, the sample was transferred to the plasma chamber and cleaned with an argon and oxygen (10%) plasma mixture. This technique primarily uses active oxygen atoms to react with carbon, producing the volatile compounds CO2 and CO, and thereby eliminating contaminants [18]. The gas mixture was introduced to a pressure of $4 \times 10^{-2}$ mbar, and the RF power supply was consistently set to a power output corresponding to approximately 50 W applied for 1 h (figure 2, procedure A). A third XPS measurement (Mid-T bake + Plasma) was performed following the plasma treatment to conclude path A.

The second experimental procedure (procedure B) was applied to the second sample. This consisted of the same steps but the sample was exposed to plasma treatment before Mid-T baking and as well exposed to air after the Mid-T bake (Plasma + Mid-T bake + 48h in air). The last step reproduces conditions similar to those experienced by SRF cavities installed in accelerators. Finally, an additional XPS measurement was performed after applying a final plasma treatment with the same gas mixture and exposure time (Plasma + Mid-T bake +48h in air + Plasma).

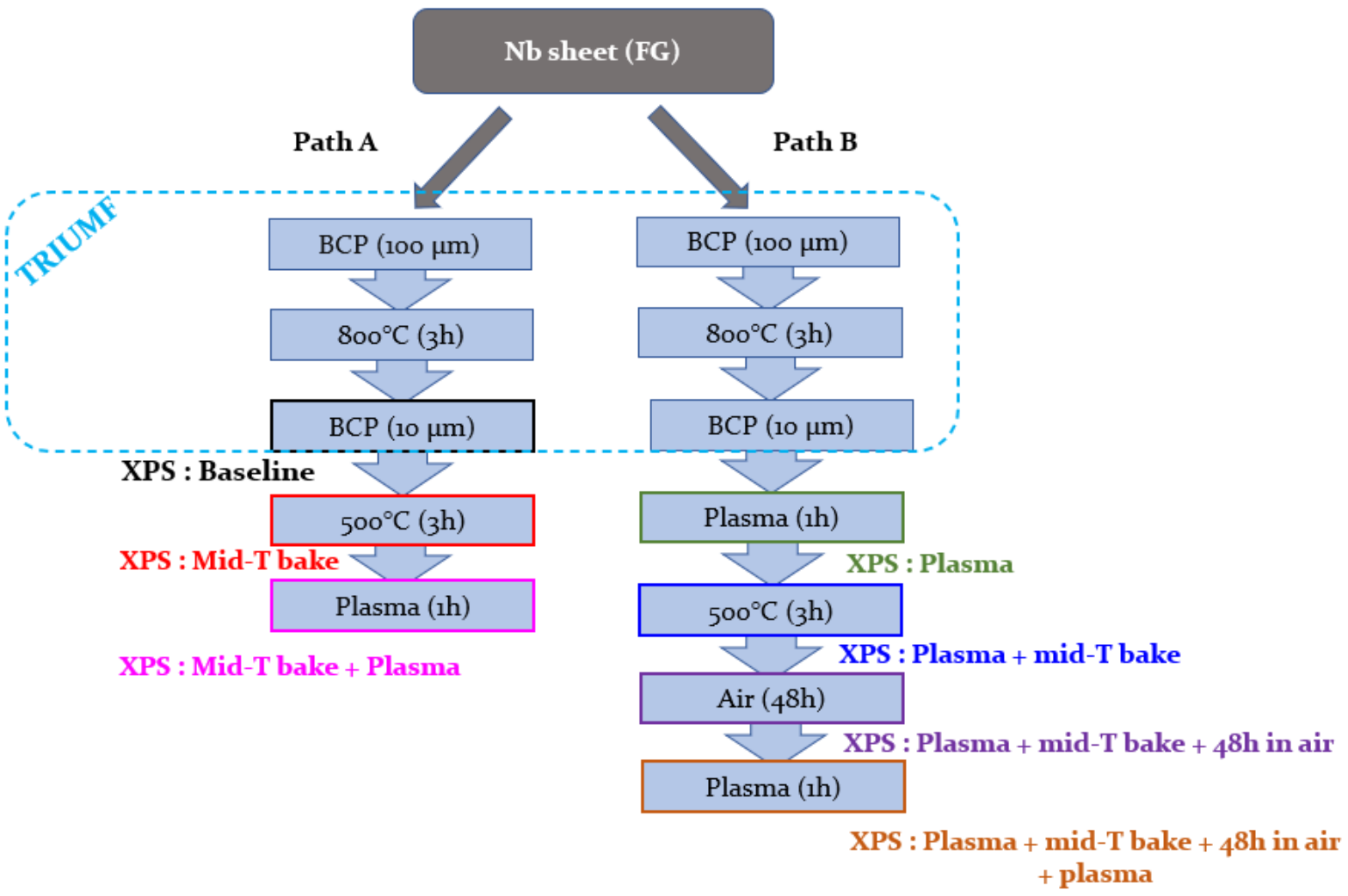

Figure. 2: List of steps for the two different experimental procedures

### *1.3 X-Ray photoelectron spectroscopy parameters*

An Al Kα x-ray source (hν = 1487 eV) with 300 µm diameter spot size and fixed at an angle of 72° from the sample normal, was used to acquire the XPS spectra. Survey spectra as shown in Figure. 3 for the baseline measurement were collected using a pass energy of 20 eV with a dwell time of 100 ms and a step size of 1 eV. High-resolution spectra were collected with a 50 eV pass energy, 50 ms dwell time and sufficient sweeps for a better signal-to-noise ratio. It should be noted that the Nb sample was exposed to ambient atmosphere, and its surface is therefore covered with carbon contamination prior to plasma cleaning reactions. To confirm this, a complete photoelectron energy distribution was measured to identify the main compounds present on the surface. After identifying Niobium and its compounds with oxygen and carbon, the peak shapes were analyzed to deconvolute the different spectral components arising from the various chemical compounds.

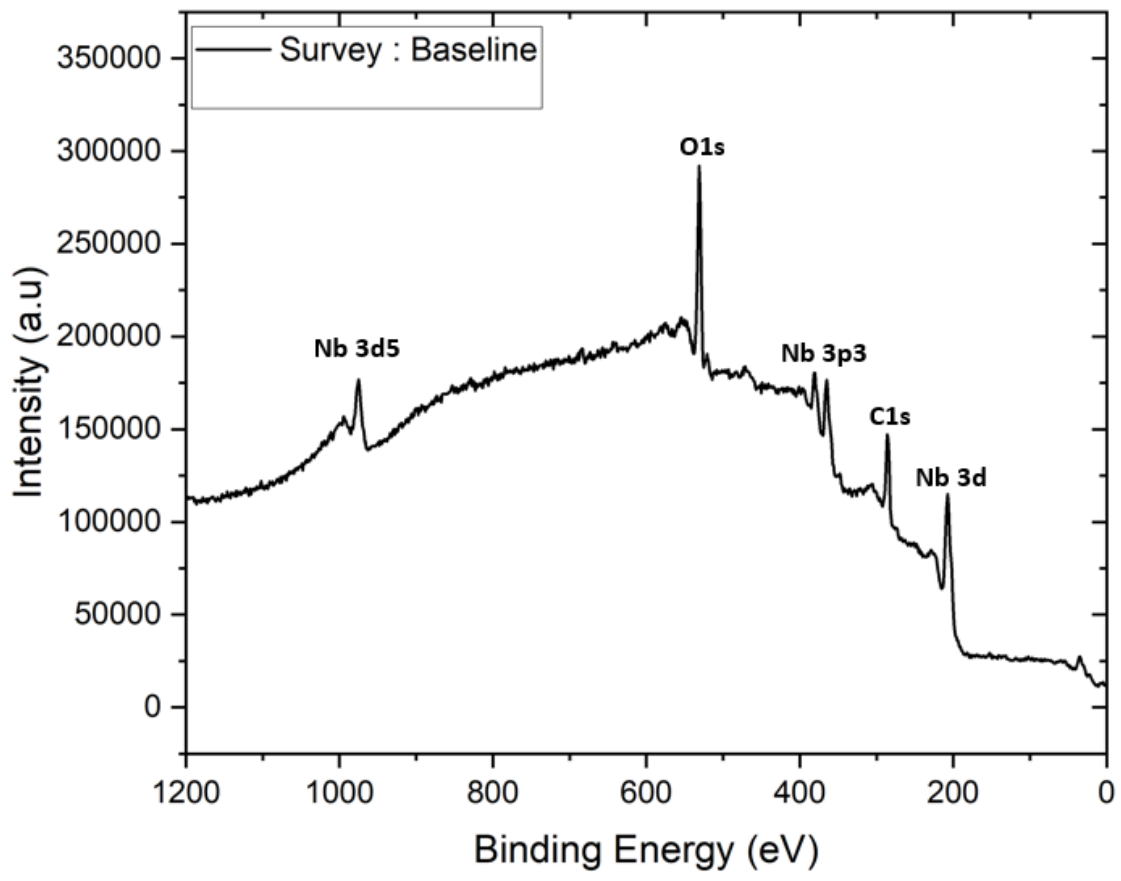

Figure 3 : XPS survey spectrum of the baseline niobium sample

The XPS spectra were deconvoluted and treated with Casa XPS software. Deconvolution of the C 1s (Figure 4 and 5) and Nb3d (Figure 10) peaks was performed using Gaussian–Lorentzian curves for the oxide and asymmetric Lorentzian curves for the metal. Background calculation was performed according to the Shirley method [14]. The FWHM of all peaks were basically smaller than 2 eV. Doublets of Nb3d3/2 to Nb3d5/2 were set apart for 2.70 eV and the ratio of corresponding peak area was set to be 2/3. In this study, the analysis was limited to the outermost surface of the samples, corresponding to an X-ray photoelectron escape depth of approximately 5 to 10 nm [31]. To obtain compositional information from deeper regions, ion sputtering using an ion gun was considered. However, preliminary investigations showed that ion bombardment at an energy of 2 keV induces carbide formation at the surface, thereby altering the initial chemical composition and affecting the analysis results (see Appendix 1).

## 2. Results

### *2.1 Influence of plasma treatment on the contamination during the Mid-T baking*

For the baseline, the C1s spectrum is characteristic of the typical adventitious carbon contamination commonly observed on metallic surfaces (figure 4(a)). It exhibits the three main contributions generally associated with atmospheric exposure. The dominant component, centred at 285.0 eV, corresponds to C–C bonds and accounts for about 91% of the total carbon signal. This peak is widely reported as the main signature of adventitious carbon on air-exposed metal surfaces [16,17]. Secondary components located at approximately 287.7 eV (C=O) and 289.0 eV (O–C=O) reveal the presence of oxygen-containing carbon species originating from atmospheric contamination. After annealing at 500 °C, the C1s spectrum shows a component at 284.3 eV attributed to graphitic carbon and another component at 282.3 eV, which corresponds to Niobium carbide (figure 4(b)). The carbide component represents approximately 69% of the total carbon signal, compared to 31% for the graphitic carbon contribution. The comparison of C1s for the path A is presented in the figure 4(c). The XPS performed after plasma treatment (Mid-T + plasma) reveals a surface completely devoid of graphitic carbon and also carbides (figure 4 (c)).

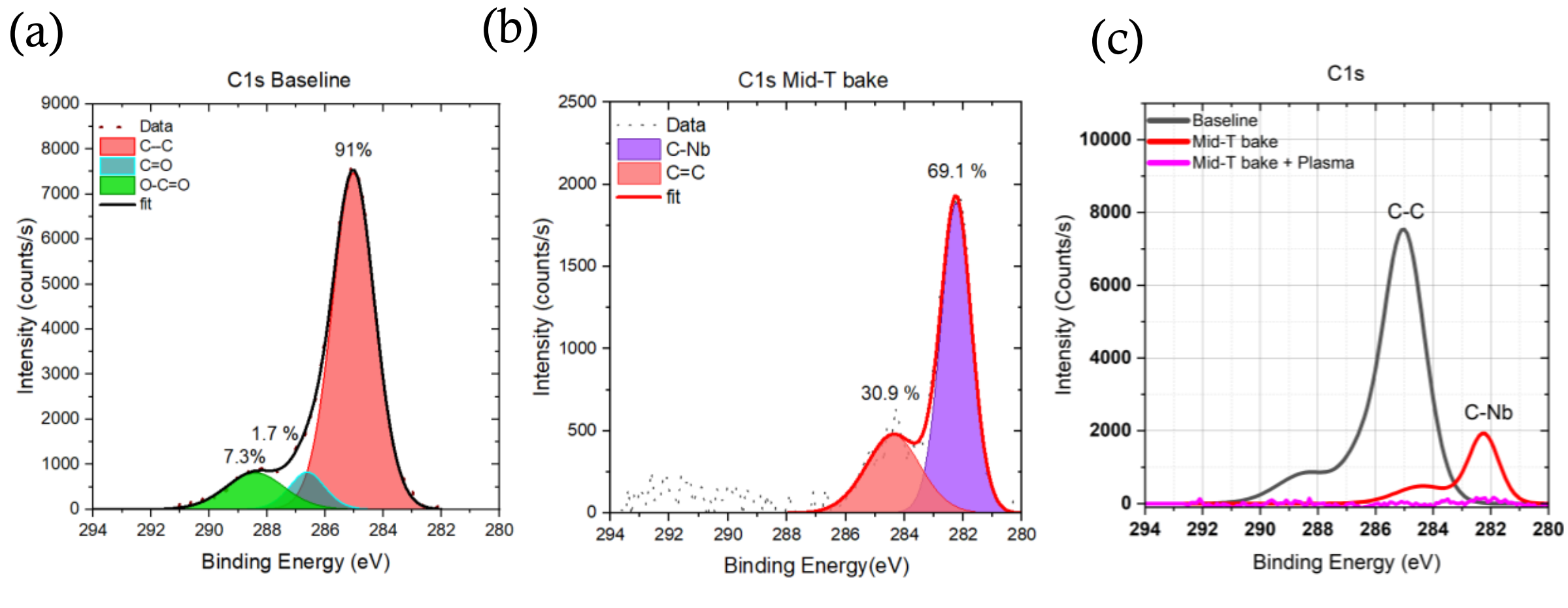

Figure 4: High-resolution XPS C1s spectra with peak deconvolution for path A: (a) baseline fitting; (b) Mid-T bake fitting; (c) comparison of the C1s peak fits for baseline, Mid-T bake, and plasma + Mid-T bake treatments.

In order to quantify to which extent the contamination layer contributes to carbide formation, the sample was treated by $Ar/O_2$ plasma before heat treatment (procedure B). Figure 5(a) shows the C1s spectra after each step-of procedure B, as outlined earlier. Measurements after plasma processing clearly show an effective removal of carbon contamination, including the C–C, C=O, and O–C=O components. After the plasma treatment and the XPS measurement, the sample was directly annealed at 500 °C for 3 h. The C1s spectrum reveals the presence of a graphitic carbon at 284.2 eV [14], as well as a Niobium carbide at 281.9 eV (figure 5(b)). The Niobium carbide component represents approximately 65.5% of the total carbon signal, compared with 34.5% attributed to graphitic carbon following annealing at 500°C for 3 hours. A summary of the contamination for all the XPS measurements in procedures A and B is given in Table 1.

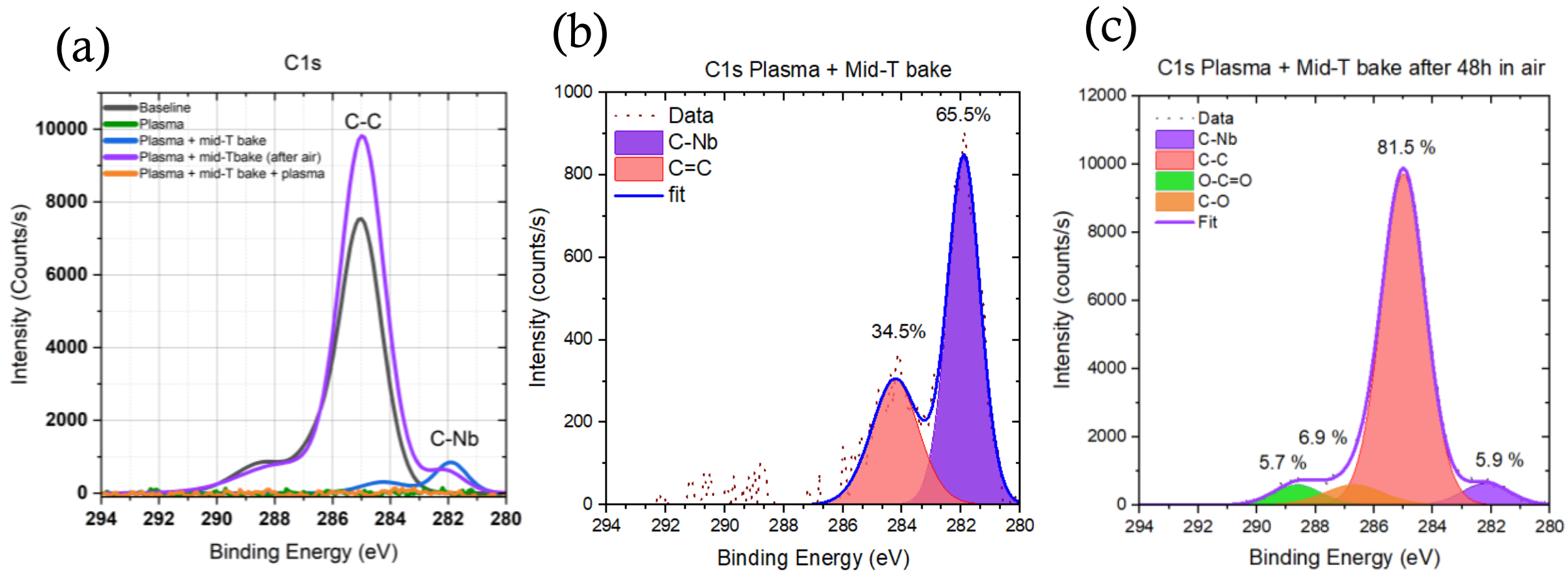

Figure 5: High-resolution XPS C1s spectra with peak deconvolution for B path: (a) baseline, plasma, plasma + Mid-T bake; plasma + Mid-T bake after 48 h in air, plasma + Mid-T bake after 48 h air exposure + plasma; (b) Plasma + Mid-T bake fitting; (c) plasma + Mid-T bake after 48 h air exposure.

Dacca et al. [15], reported that the amount of carbon present on the surface after a thermal cycle is consistent with the growth of niobium carbide and graphite as observed in the C1s spectrum (66.6 % of niobium carbide and 33.3 % at of graphite). They attributed this contamination to only the residual gases in the analysis chamber, mainly CO2 and CO despite the vacuum of $10^{-9}$ mbar during the baking. Surface contamination is one of the main sources of carbon feeding carbide formation. Residual gas and carbon bulk diffusion during chemical etching seem to be two other equivalent sources of carbon according to the studies conducted by Septier in 1980 [1].

The XPS spectrum after exposure to air (figure 5(c)) indicates the presence of carbon contamination (C–C) at 285.0 eV, C-O at 286.6 eV, O–C=O at 288.6 eV and C-Nb at 282.2 eV. The application of a plasma treatment prior to annealing made it possible to eliminate the C=O component and to promote the formation of a C–O component (see table 1).

Finally, the plasma treatment of the sample after air exposure fully removed niobium carbide on the surface (orange curve in figure 5(a)), [19]. This observation is particularly interesting for the application on SRF cavities. Indeed, reactive plasma decontamination is now widely applied on cavities as recovery treatment to mitigate field emission [28]. This could be now applied as preventive treatment to purify the surface and potentially mitigate side effects induced by Mid-T baking namely the High Field Q-Slope and the early quench field [6] or the higher residual magnetic field sensitivity [27].

| Treatments | C species | Positions (eV) | Quantifications (%) |
|---|---|---|---|
| Baseline (A and B) | C-C | 285.0 | 91.0 |
| | C=O | 287.7 | 1.7 |
| | O-C=O | 289.0 | 7.3 |
| | C-Nb | – | – |
| Mid-T bake (A) | C=C (graphite) | 284.3 | 30.9 |
| | C-Nb | 282.3 | 69.11 |
| Mid-T bake + Plasma (A) | C=C (graphite) | – | – |
| | C-Nb | – | – |
| Plasma (B) | C-C | – | – |
| Plasma + Mid-T bake (B) | C-C | 284.2 | 34.5 |
| | C-Nb | 281.9 | 65.5 |
| Plasma + Mid-T bake after air (B) | C-C | 285.0 | 81.5 |
| | C-O | 286.6 | 6.9 |
| | O-C=O (graphite) | 288.6 | 5.7 |
| | C-Nb | 282.2 | 5.9 |
| Plasma + Mid-T bake + Plasma (B) | C-C | – | – |
| | C-Nb | – | – |

Table 1: Summary the binding energies and relative contributions of the C1s XPS components for the different niobium surface treatments.

## *2.2 Identification of $Nb_2C$ and NbC carbide phases*

Figure 6 shows the C1s XPS spectra of Niobium samples after annealing (a) and after annealing preceded by a plasma treatment also called “Plasma Bake or Plasma + Mid-T bake” (b). The comparison of areas under the curves of C-Nb for Mid-T bake (Area = 2231) and the Plasma + Mid-T bake (Area = 1049) treatments show that the plasma bake reduces carbide formation by 53% compared to Mid-T alone. Interestingly, associated to the reduction in the intensity of the carbide component observed in figure 4(c) and 5(a), is a shift in binding energy of approximately 0.4 eV as shown in figure 6. The measurements in figure 6 (a) indicate that annealing at 500°C alone results in a carbide component at 282.3 eV, which corresponds to $Nb_2C$ [22]. In contrast, the spectrum of the sample that underwent plasma treatment before annealing shows in figure 6(b) a contribution at 281.9 eV, which corresponds to NbC [32]. This energy shift could be clearly observed thanks to the very clean surface conditions with very low carbon contamination, which could otherwise mask the signal. For an air exposed sample, figure 5(c), contamination becomes dominant: the NbC binding energy, measured *in-situ* at 281.9 eV, shifts to 282.2 eV in the *ex-situ* measurement (Table 1).

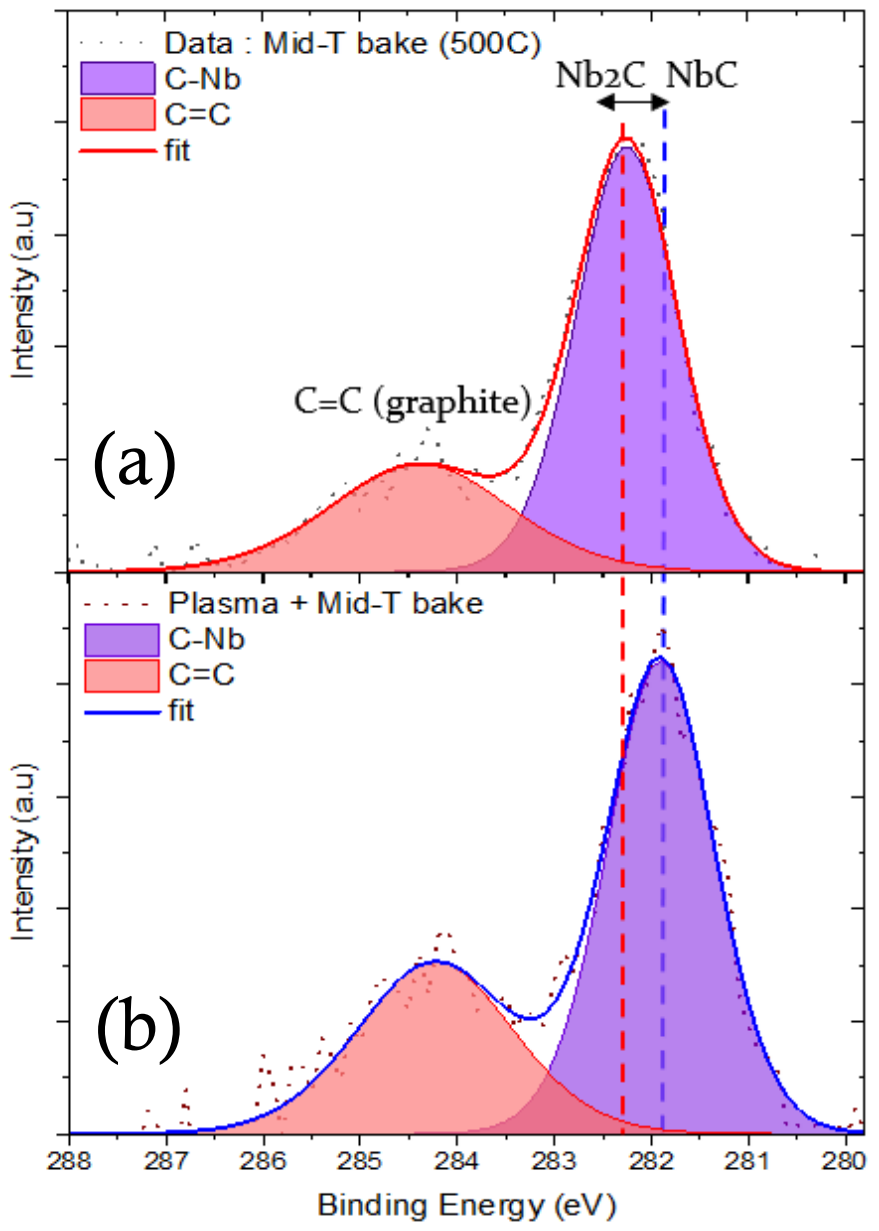

Figure 6: High-resolution XPS C1s spectra showing $Nb_2C$ formation after Mid-T bake (a) and NbC formation after plasma + Mid-T bake (b).

In an effort to clearly identify the two phases, the samples have been analyzed by ex-situ SEM after air exposure. The surface morphology information for the initial state (Baseline), Mid-T bake, plasma treatment followed by a baking, and final plasma treatment are shown in figure 7 obtained by SEM in a secondary-electron mode. The baseline, figure 7(a), free of carbides, shows the Niobium surface without any precipitates. The image in figure 7(b) demonstrates the submicrometer-size pyramid-shape precipitates which are assigned to a Nb2C phase [22]. Fig. 7(c) shows that a plasma treatment after annealing is effective at removing $Nb_2C$ surface structures. For the procedure B, the Nb sample with plasma followed by annealing, figure 7(d), fine white spots are observed, which are characteristic of the NbC phase. No contamination is observed after the last plasma processing, figure 7(e) [13].

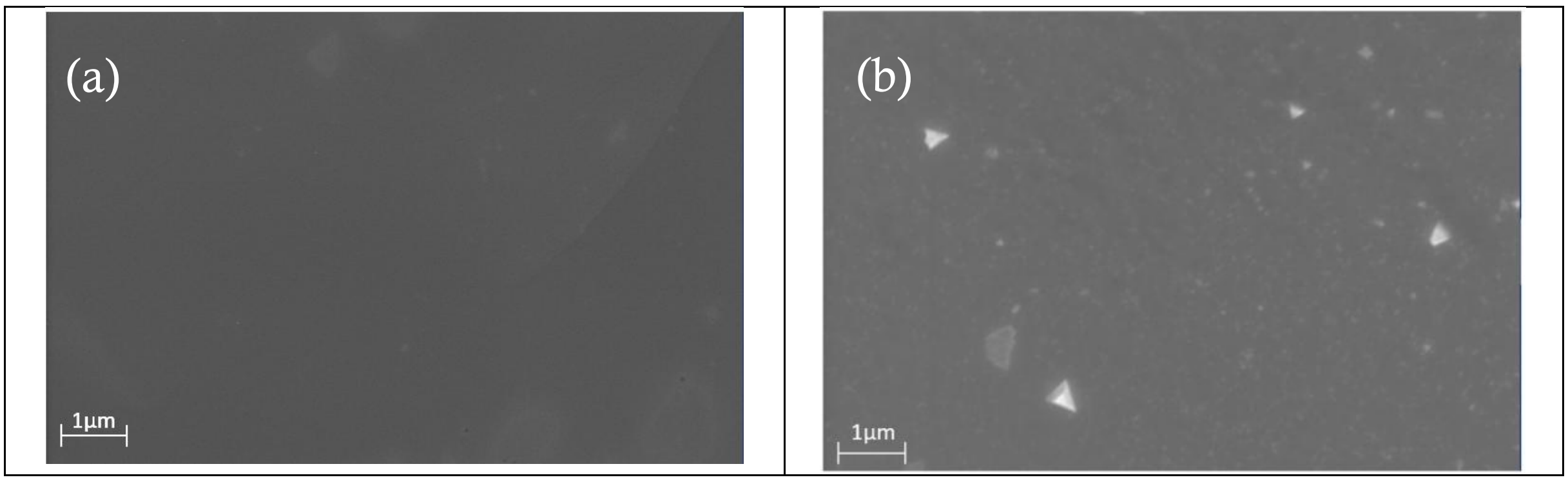

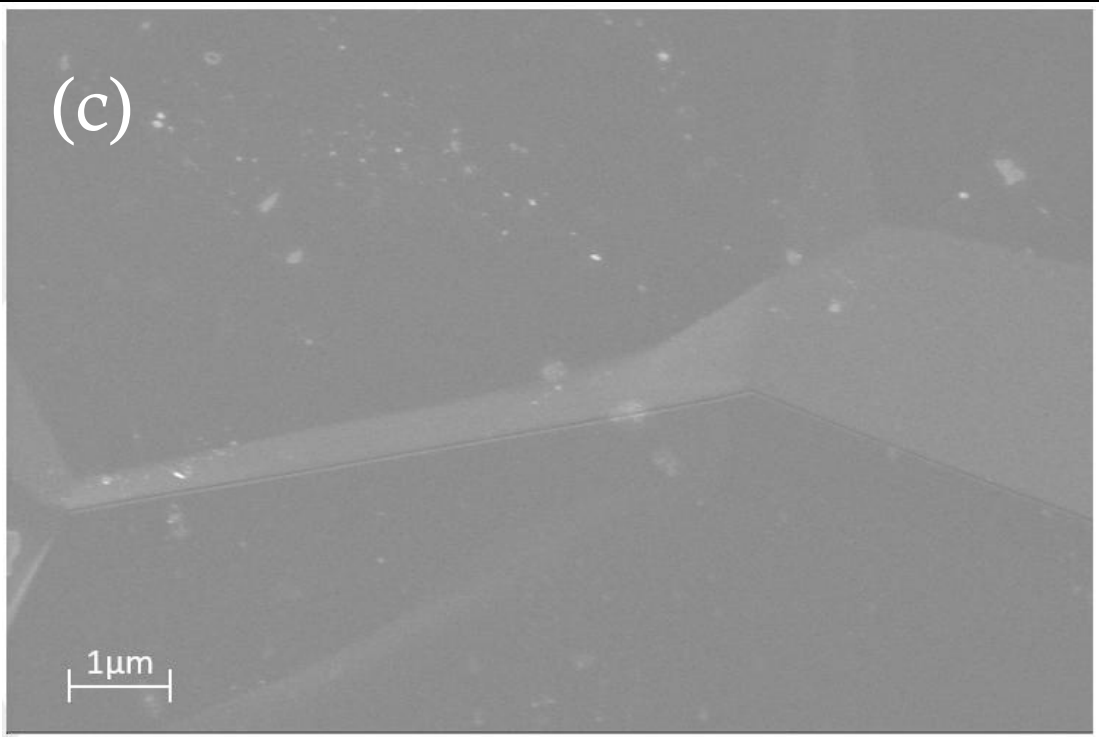

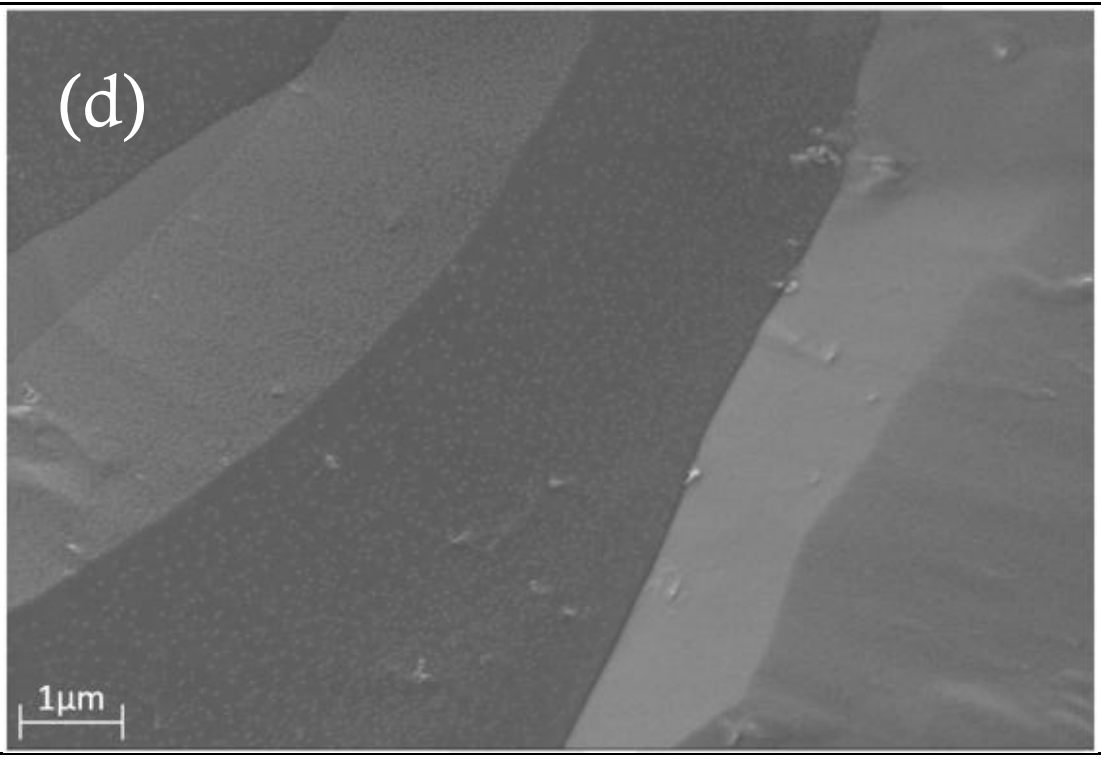

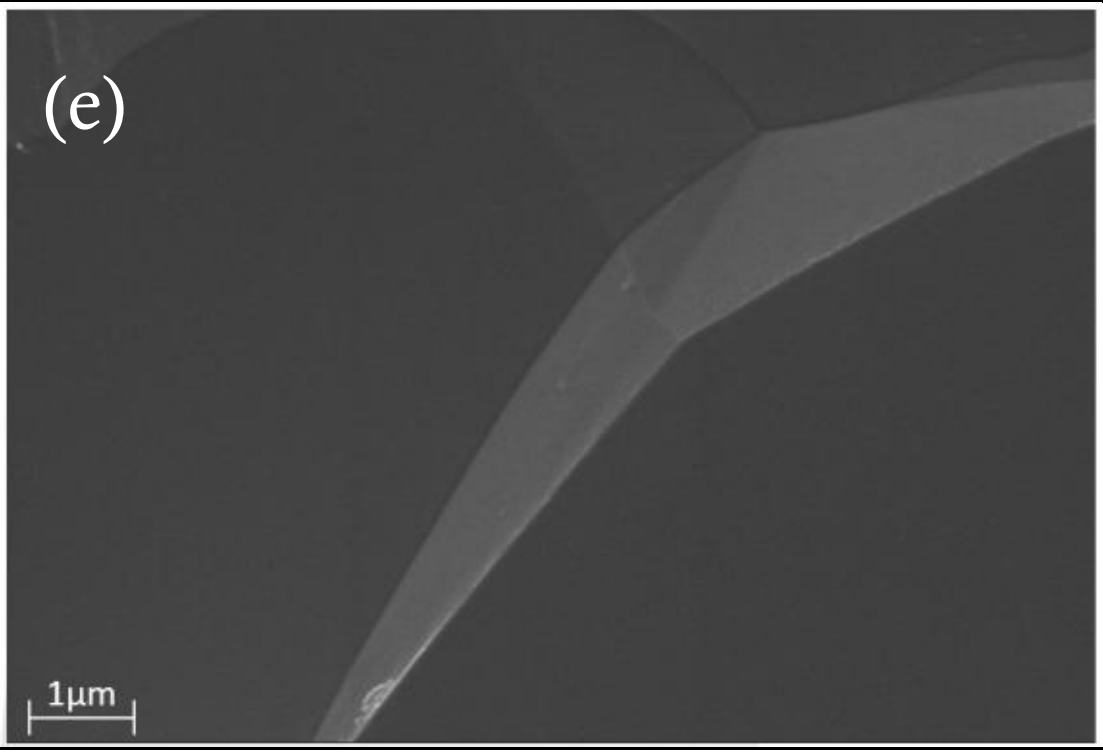

Figure 7: SEM (Secondary electron mode) images of the niobium surface for different treatments: (a) baseline; (b) Mid-T bake; (c) Mid-T bake + plasma; (d) plasma + mid-T bake; and (e) plasma + Mid-T bake + plasma.

### *2.3 Modification of surface oxide during plasma treatment*

By studying the impact of plasma cleaning on the formation of Niobium carbides, interesting observations could be done on the oxide layer. The native Niobium oxide, typically of 5-7 nm, is composed of several layers with an increasing degree of oxidation for the outermost layers. Starting at the bulk and moving to the surface the layers are NbO, $NbO_2$ and $Nb_2O_5$ is the only oxide stable in atmosphere [23]. The spectra of O1s in figure 8 and Nb3d in figure 9 were obtained in the binding energy range of 536 -526 eV and 212–197 eV, respectively. The Baseline exhibited a significant amount of oxide on the surface. The signal recorded at 530.4 eV corresponds to $Nb_2O_5$ and is also visible in the Nb 3d spectrum at 207.0 eV for the 5/2 orbital and 209.7 eV for the 3/2 orbital. The metallic doublets in figure 9 (a-b) are also visible at 201.7 eV and 204.5 eV. After plasma treatment, the O1s in figure 8 (a-b) and Nb3d in figure 9 (a-b) show a clear energy shift, respectively from 530.4 eV to 529.5 eV and from 207.0 eV to 206.1 eV, corresponding to a reduction of the $Nb_2O_5$ into $NbO_2$ oxide [17]. This shift is also observed each time the sample was directly exposed after 1 hour to the plasma cleaning. This reduction can be attributed to a preferential sputtering mechanism rather than a chemical reduction process. Ion energy distribution studies [33] performed with a Langmuir probe on a similar plasma source have shown that in an $Ar/O_2$ mixture, $O^+$, $O_2^+$ and $Ar^+$ ions reach bombardment energies of 40–61 eV, compared to only ~2 eV for $Ar^0$ in a pure Ar plasma, due to kinetic coupling between Ar and oxygen ions [33]. When the resulting sputtering rate exceeds the oxide regrowth rate, the $Nb_2O_5$ surface layer is progressively removed, exposing the underlying $NbO_2$ layer. This reduction was previously observed by D. K. Murti & R. Kelly [34] in 1976. They showed that ion bombardment of $Nb_2O_5$ preferentially sputters oxygen, leading to oxygen-deficient regions [34-35] and the formation of lower-stoichiometry oxides such as $NbO_x$/NbO [34].

After Mid-T baking, the oxide is significantly dissolved but still visible in the O1s spectra of figure 8. Interestingly, in the case of the sample following procedure B, which was treated by plasma prior to baking, the oxide ($NbO_2$) is not only dissolved but the surface became more metallic, which is reflected by a shift to the right (see figure 9 (b)). After exposure to air, the $Nb_2O_5$ oxide layer reforms as visible in figure 8(b) and 9(b). However, the re-oxidation tends to grow a thinner oxide (dry process) compared to the baseline (wet process after BCP chemical etching) as the amplitude of metal doublets is significantly higher.

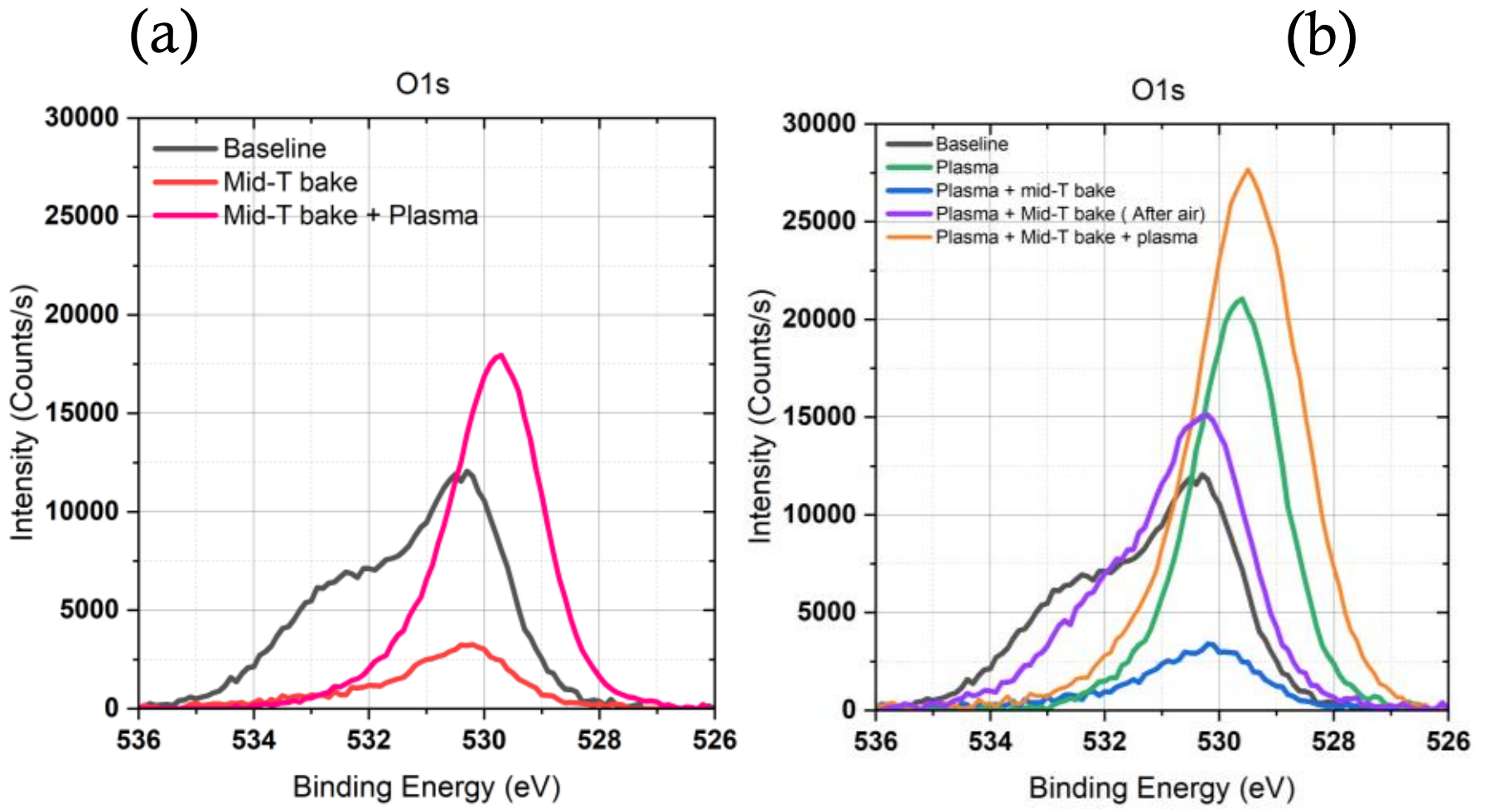

Figure 8: (a) Procedure A: Comparison of the O1s XPS peak fits for baseline, Mid-T bake, and Mid-T bake+ Plasma; (b) Procedure B: Comparison of the O1s XPS peak fits for baseline, plasma, plasma + Mid-T bake; plasma + Mid-T bake after 48 h air exposure, plasma + Mid-T bake after 48 h air exposure + plasma

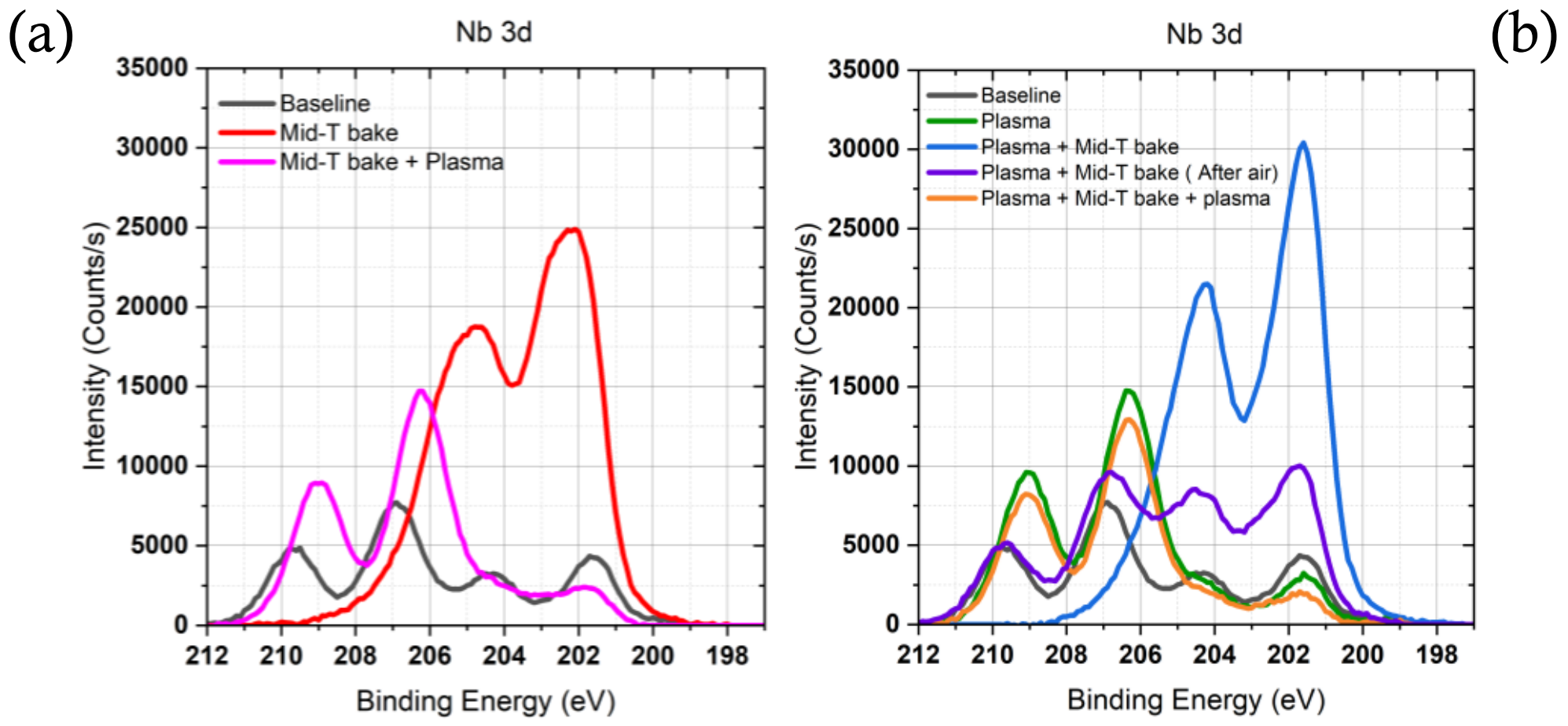

Figure 9: (a) Procedure A: Comparison of the O1s XPS peak fits for baseline, Mid-T bake, and Mid-T bake+ Plasma; (b) Procedure B: Comparison of the O1s XPS peak fits for baseline, plasma, plasma + Mid-T bake; plasma + Mid-T bake after 48 h air exposure, plasma + Mid-T bake after 48 h air exposure + plasma

Finally, the changes in stoichiometry of the Niobium oxide layers by the various treatments have been analyzed by deconvolution of the Nb3d XPS spectra and are presented in figure 10(a) for Mid-T bake treatment (procedure A) and in figure 10(b) for Plasma + Mid-Bake treatment (procedure B). A clear difference in doublet peak shapes is visible meaning that the plasma treatment prior to heat treatment has a non negligeable impact on the surface stochiometry of Niobium. For the sample following procedure A, the deconvolution of the Nb 3d XPS spectrum (figure 10(a)) and quantification reveals a NbO phase, accounting for about 28.8 % at 202.8 eV for the 5/2 orbital, a sub-oxide $Nb_xO$ ($1 < x < 2$), representing ≈28 % at 202.4 eV. Metallic niobium (Nb°) contributes to about 27.2 % at 201.7 eV, while the carbide fraction ($Nb_2C$) represents ≈ 16 % at 204.0 eV. For the sample following procedure B, when plasma treatment is applied prior to annealing (figure 10(b)), the surface becomes significantly more metallic. The metal proportion increases to ≈ 70.5 %, whereas no $Nb_xO$ phase is visible. Simultaneously, the NbO fraction decreases to ≈20 %, and the NbC contribution is reduced to ≈ 9.5%. This improvement of the Nb surface will be called "Plasma Bake". These results in Table 2 indicate that plasma treatment efficiently removes unstable oxide and carbon contaminants, leading to a cleaner and more metallic niobium surface after heat treatment.

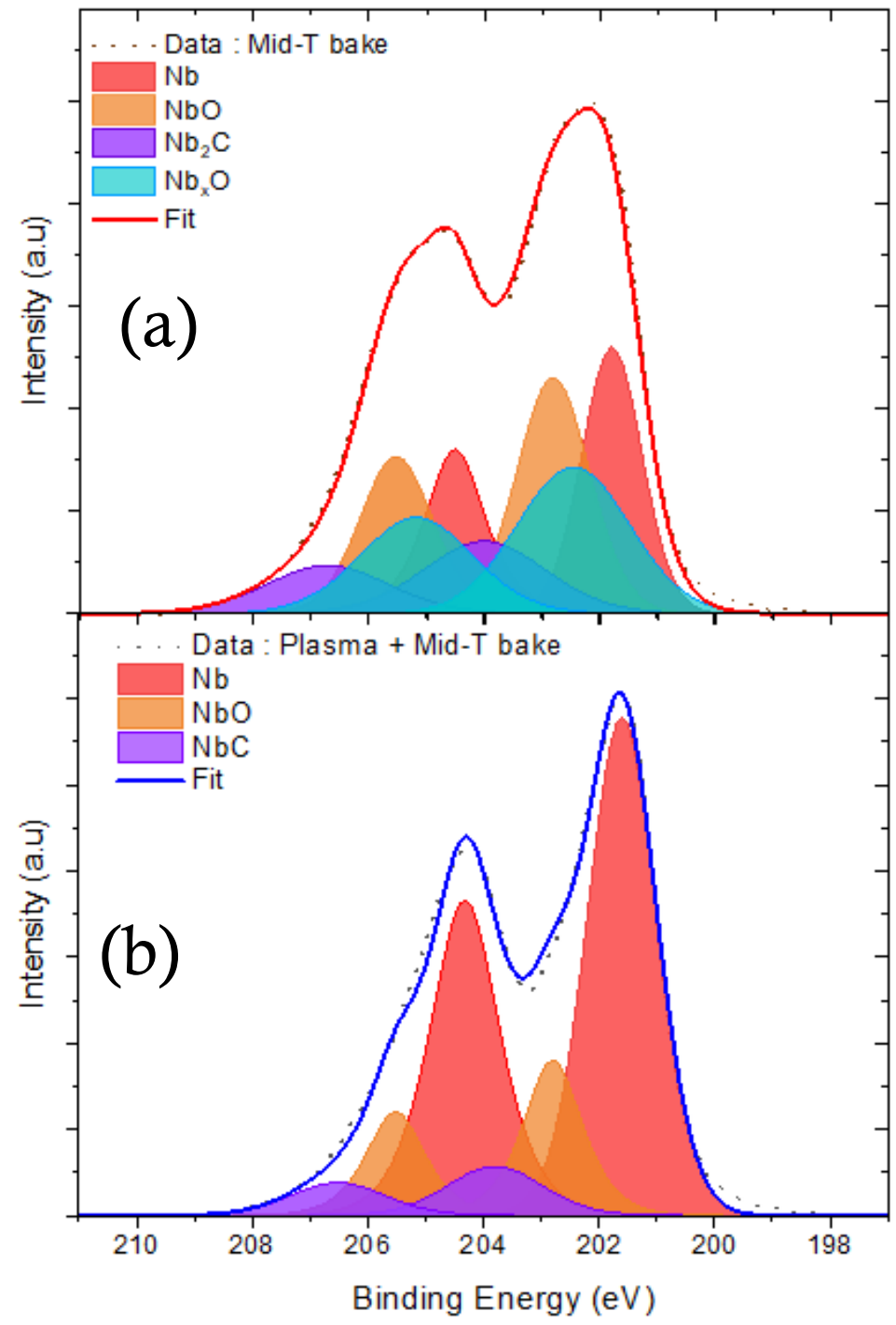

Figure 10: High-resolution XPS N3d spectra of Mid-T bake (a); Plasma + Mid-T bake (b).

| Treatments | Nb species | Positions (eV) | Quantifications (%) |
|---|---|---|---|
| Mid-T Bake (A) | Nb3d (5/2 and 3/2) | 201.7<br>204.4 | 27.2 |
| | NbO | 202.8<br>205.5 | 28.8 |
| | $Nb_2C$ | 204.0<br>206.7 | 16 |
| | $Nb_xO$, $1 < x < 2$ | 202.4<br>205.1 | 28 |
| Plasma + Mid-T bake (B) | Nb3d (5/2 and 3/2) | 201.6<br>204.3 | 70.5 |
| | NbO | 202.8<br>205.5 | 20 |
| | NbC | 203.8<br>206.5 | 9.5 |

Table 2 Binding energies and relative contributions of the Nb3d XPS components for the different niobium surface treatments.

The major processes occurring at the surface of the sample during the thermal treatments can be summarized by the transition $Nb_2O_5$ to NbO at T=280-500°C. The transition from $Nb_2O_5$ to NbO involves a progressive reduction of the niobium oxidation state ($Nb_2O_5 \rightarrow NbO_2 \rightarrow Nb_xO \rightarrow NbO$) [15][20]. Now, the reduction of oxides by "Plasma Bake" is given $NbO_2 \rightarrow NbO$. This reduction occurs simultaneously with a change in the chemical state of carbon, which evolves from graphitic carbon to Nb-C bonding [15].

## 3. Discussion on implication for SRF applications

We have discovered that applying *in-situ* plasma treatment before thermal annealing reduces carbide formation by approximately 53%, and also modifies the chemical nature of the carbides. The plasma treatment promotes the formation of the superconducting NbC phase rather than $Nb_2C$, the latter being intrinsically non-superconducting [21]. In addition, the plasma treatment with 10% of oxygen modifies a $Nb_2O_5$ layer into $NbO_2$. During ultra-high vacuum baking, plasma processing allowed elimination of unstable sub-oxides such as $Nb_xO$, while the proportion of metallic niobium at the surface significantly increases, resulting in a cleaner and more metallic surface state after heat treatment. This is particularly interesting as a potential method to improve the efficiency of Mid-T baking processes applied to SRF cavities. Doping recipes in SRF application have been shown to provide enhanced quality factor (reduced RF surface resistance) as a function of applied RF field, also called anti-Q slope. Carbide formation above certain temperatures have been shown to limit the achievable $Q_0$. Plasma treatment has the potential to further optimize mid-T baking recipes in terms of enhancing the achievable $Q_0$ and gradient. From our study, the growth of Niobium carbide ($Nb_2C$ or NbC) above 300°C seems to be unavoidable even in the best heat treatment conditions practically achievable on large UHV furnaces dedicated to SRF cavities (typically in the $10^{-7}$ to $10^{-8}$ mbar). We have shown that carbides are formed even on a Niobium sample free of carbon contamination and baked in a vacuum in the range of $10^{-9}$ mbar. Previous studies [24], have shown that the presence of the $Nb_2C$ phase is detrimental for superconducting properties [20] leading to a reduction in the anti-Q slope, an indication of the degradation of the surface resistance versus accelerating gradient. On the other hand, NbC phase, obtained in low carbon contamination environment, because of its superconducting properties, may not be problematic, or may, on the contrary, participate to the improvement of $Q_0$ factor. This improvement of $Q_0$ is today only attributed to the diffusion of oxygen impurities (oxygen doping) [6]. More specifically for the Q-bit application of SRF cavities, the plasma treatment prior to heat treatment is particularly interesting as the final surface is more metallic. Indeed, the $Q_0$ factor is significantly altered by the oxide layer in the very low field region, as dielectric losses are dominating [25].

As the application of a plasma treatment prior to heat treatment without exposing the cavity walls to atmosphere might not be straightforward, we have shown as well that a plasma treatment applied after annealing eliminates efficiently carbides (NbC and $Nb_2C$) but on the other hand increases the oxide thickness. This observation is particularly interesting for the application on SRF cavities. Indeed, reactive plasma decontamination is now widely applied on cavities as a recovery treatment to mitigate field emission [26]. Plasma treatment could now be applied as a preventive treatment to remove carbide contamination from the Niobium surface and potentially mitigate side effects induced by Mid-T baking namely the High Field Q-Slope, the early quench field [6] or the higher residual magnetic field sensitivity [27]. Preliminary tests have shown in Figure 11 (a-b) that pure Argon plasma can as well remove C–C, C=O, and O–C=O components in Figure 11(a), however with lower efficiency without converting a $Nb_2O_5$ layer into $NbO_2$. This preventive $Ar^+$ plasma treatment proves that the 10% oxygen is responsible for the conversion of $Nb_2O_5$ into $NbO_2$.

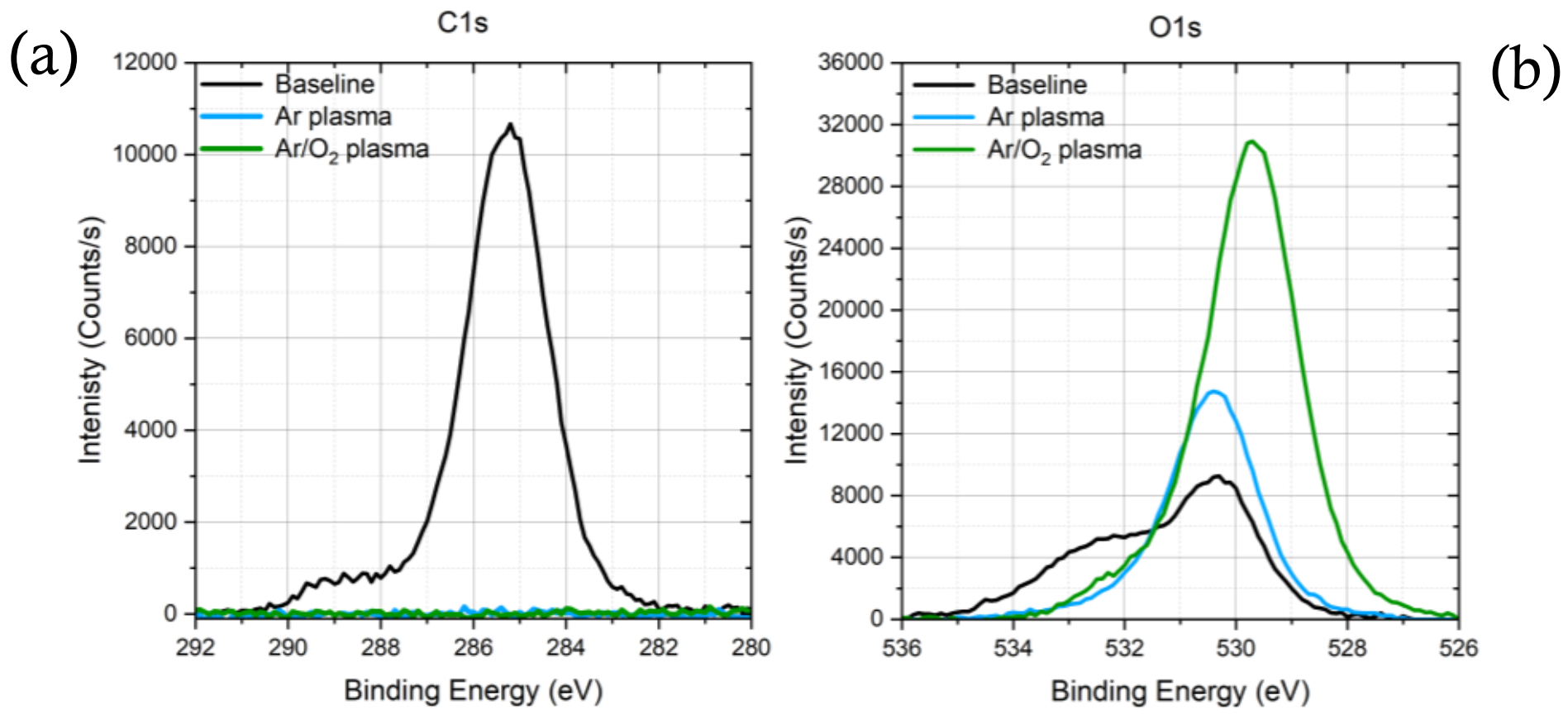

Figure 11: Comparison of the C1s XPS peak fits for baseline and Ar Plasma (a); Comparison of the O1s XPS peak fits for baseline and Ar Plasma (b).

## Conclusions

This study highlights the effectiveness of *in-situ* $Ar/O_2$ plasma treatment for improving the surface condition of bulk niobium treated by buffered chemical polishing. XPS and SEM analyses show that an $Ar/O_2$ plasma efficiently removes surface organic contamination, particularly hydrocarbons, thereby limiting the formation of niobium carbides after heat treatment that are detrimental to SRF cavity performance. This study has shown that the removal of carbon contamination and the thickening of the oxide layer prior to heat treatment due to a reactive plasma treatment promotes the formation of NbC carbides instead of $Nb_2C$. The NbC carbides are viewed as favorable in terms of superconducting properties compared to Nb2C. Plasma treatment after heat treatment and air exposure, can remove efficiently carbides of both phases. This study has paved the road to future works, which could investigate whether different plasma gas mixtures can further modify the electronic properties of the surface, particularly the secondary electron emission yield (SEY). The final step is to test the plasma bake on a cavity at TRIUMF.

## CRediT authorship contribution statement

**C.Boutelaa**: Writing – original draft, Investigation, Data Curation, Formal analysis, Visualization, Methodology, Conceptualization, Resources. **S.Gruszka**: Visualization, Methodology, Conceptualization. **C.Cheney:** Resources. **T.Gerardin**: Resources. **B.Mercier**: Resources, supervision. **E.Mistretta**: Resources. **J.Demailly**: Resources. **J.Yemane**: Resources. **S.Guilet:** review and Methodology**. R.Laxdal**: review & editing, Validation, Supervision, Resources, Methodology. **P. Kolb**: review & editing, Validation, Supervision, Resources. **J. Keir**: Resources. N.**Prud'homme**: Review & editing, Validation, Supervision. **G.Sattonnay**: Review, Validation, Supervision, Resources, Project administration, Methodology. **D.Longuevergne**: Writing – review & editing, Validation, Supervision, Resources, Project administration, Methodology, Funding acquisition, Conceptualization.

## Declaration of competing interest

The authors declare that they have no known competing financial interests or personal relationships that could have appeared to influence the work reported in this paper.

## Acknowledgment

This work was supported by CNRS Nuclei&Particles Institute and the CNRS-TRIUMF International Research Laboratory NPAT (Nuclear Physics, Nuclear Astrophysics and Accelerator Technology). This work was as well supported by Ile de France Region under grant agreement SESAME EX039203. The authors would like to thank the members of the platform "Vide&Surfaces" at IJCLab and the SRF team at TRIUMF for the technical support and the preparation of Nb samples.

## ORCID iDs

**C.Boutelaa**: https://orcid.org/0009-0009-2206-8579
**C.Cheney**: https://orcid.org/0009-0005-5536-0243
**D.Longuevergne :** https://orcid.org/0000-0003-3725-9473
**G.Sattonnay:** https://orcid.org/ 0000-0001-9759-0577

## Appendix 1

It is currently difficult to determine whether carbide removal occurs only at the surface or also extends deeper into the material. To address this issue, sputtering using an ion gun under $Ar^+$ bombardment would be required. *In-situ* sputtering was performed at 2 keV for 10 min directly on the Nb sample in order to analyze deeper regions (figure 1(a-b)). Figure 1(a) presents the C1s XPS peak before and after abrasion. Surface contamination (C–C) was removed but the contribution has observed in the figure 1(a-b). The peak positions shown in figure 1(b) reveal graphite at 284.3 eV and niobium carbide at 282.3 eV, corresponding to the contribution of $Nb_2C$, which is also visible by SEM through *ex-situ* measurements in Figure 1(c). Indeed, the supplied energy is sufficient to form graphite (C=C) and niobium carbides (Nb–C) [29-30]. In our study, it will not be possible to characterize the electronic structures in depth due to the energy delivered by the ions during their collision with the Nb surface.

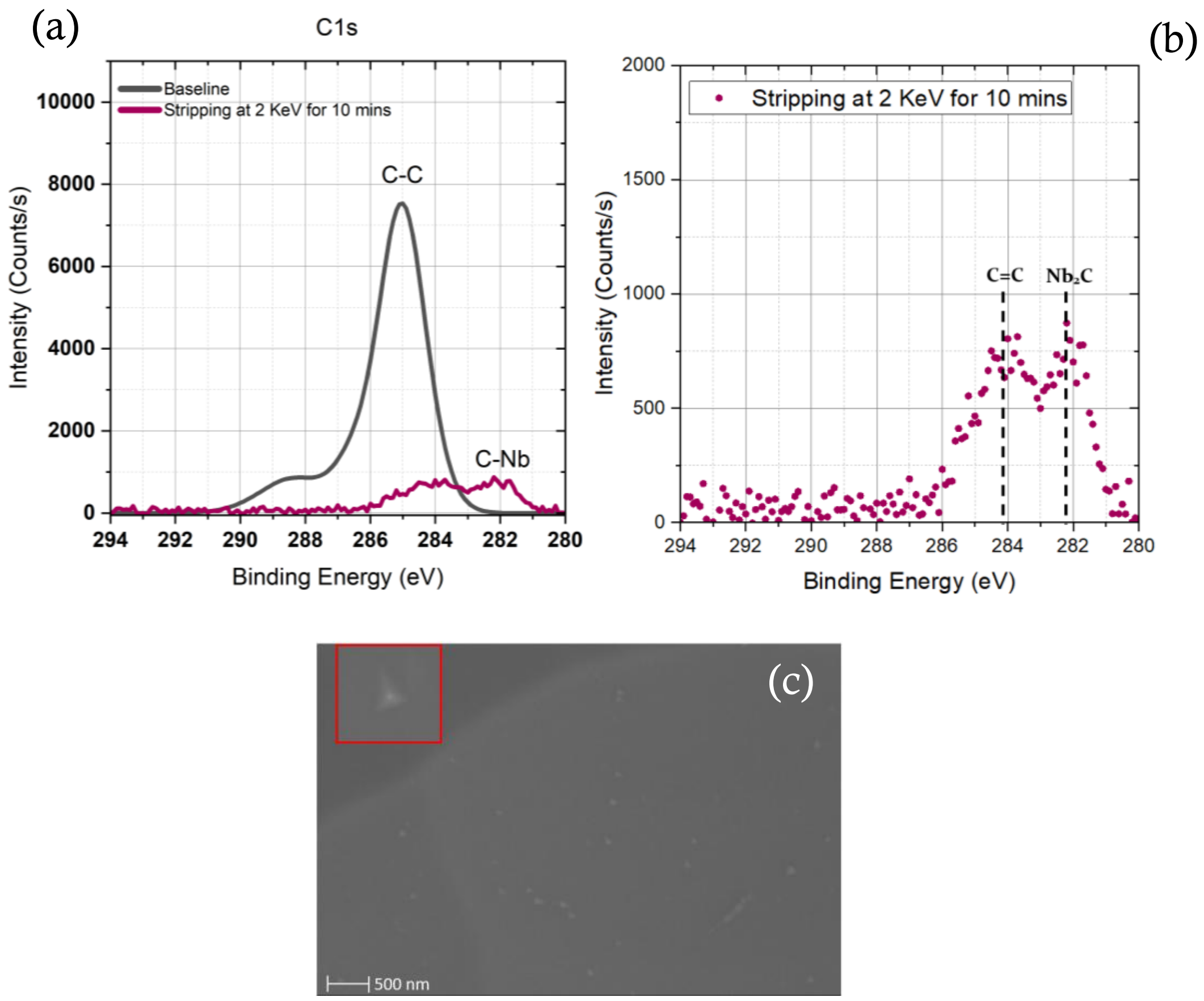

Figure 1: *In-situ* comparison of the C1s peak XPS before and after abrasion at 2 KeV for 10 mins (a) with $Ar^+$ bombardment by ion gun; Positions peaks XPS for abrasion at 2 KeV for 10 mins (b); *Ex-situ* SEM pictures after abrasion at 2 KeV (c).